\documentstyle[preprint,aps]{revtex}
\newcommand{\be}{\begin{equation}} \newcommand{\ee}{\end{equation}} 
\newcommand{\bea}{\begin{eqnarray}}\newcommand{\eea}{\end{eqnarray}}

\begin{document}
\draft
\preprint{IMSc/98/07/47, IP/BBSR/98-25, solv-int/9808005}
\title{ 
Relationship Between the Energy Eigenstates of Calogero-Sutherland Models 
With Oscillator and Coulomb-like
Potentials}
\author{Pijush K. Ghosh$^{a,1}$ and Avinash Khare$^{b,2}$}
\address{ $^a$ The Institute of
Mathematical Sciences,
CIT Campus, Taramani,
Chennai-600 113, INDIA.\\
$^b$ Institute of Physics,
Sachivalaya Marg,
Bhubaneswar-751 005, INDIA.\\}
\footnotetext{$\mbox{}^1$E-mail address: 
pijush@imsc.ernet.in }  
\footnotetext{$\mbox{}^2$E-mail address: khare@iopb.stpbh.soft.net}

\maketitle
\begin{abstract} 
We establish a simple algebraic 
relationship between the energy eigenstates 
of the rational
Calogero-Sutherland model with harmonic oscillator
and Coulomb-like potentials. We show that there is an underlying $SU(1,1)$
algebra in 
both of these models which plays a crucial role in such an identification.
Further, we show that our analysis is in fact valid for any many-particle
system in arbitrary dimensions whose potential term (apart from the oscillator
or the Coulomb-like potential) is a homogeneous function
of coordinates of degree $-2$. The explicit coordinate transformation
which maps the Coulomb-like problem to the oscillator one has also 
been determined
in some specific cases.
\end{abstract}
\narrowtext

\section{Introduction}

The rational Calogero-Sutherland model (CSM) describes a system of $N$
particles interacting with
each other via a long range inverse square interaction \cite{cs,cs1,pr}
and are confined on a line by a simple harmonic oscillator (SHO) potential.
This model is exactly
solvable and the spectrum as well as the eigen functions are well known.
Further, it is known that the rational CSM, with the SHO potential
replaced by a Coulomb-like interaction, is also exactly solvable
\cite{iop}.
The remarkable common
feature of both the models is that they reduce
to the usual harmonic oscillator or the 
Coulomb-like problem in dimensions greater than one,
once the the short distance correlations are factored out. 

It is worth pointing out that 
the only two problems, which can be solved for all partial waves in 
dimensions greater than one, are 
the usual harmonic oscillator and the Coulomb problems. Further, 
a mapping relating the energy eigenvalues 
as well as the eigenfunctions of these
two models exists in any number of dimensions\cite{china1,china2}.
It is then natural to enquire if there is a 
mapping between the energy eigenvalues as
well as eigenfunctions of the rational CSM and the same quantities of the
CSM with the Coulomb-like interaction.

The purpose of this paper is to show that such a mapping 
between these two types of CSM indeed exists. In particular, we show that
both the models posses an underlying $SU(1,1)$ algebra with different
realizations for the generators of the algebra, much akin to the
usual harmonic oscillator or the Coulomb problem \cite{china1,china2}. Using
this underlying algebra, we show that the energy eigenvalues as well as the
eigenfunctions of the rational CSM with the Coulomb-like interaction 
can be obtained from the corresponding
CSM oscillator problem. Our results are valid for all types of
rational CSM, namely, the CSM associated with the root structure of $A_N$,
$B_N$, $C_N$, $BC_N$ and $D_N$. Thus, we are able to generalize the $A_N$ type
of CSM with Coulomb-like interaction \cite{iop} to 
$BC_N$, $B_N$, $C_N$
and $D_N$ type and hence show that all these models are also exactly
solvable. Thus we are adding  
new members to the family of the exactly solvable
one dimensional many-body systems. 

We also generalize these results to several 
higher dimensional Calogero-Sutherland type of models. In particular, we
show that such a mapping is possible in any arbitrary dimension provided
the long-range many-body interaction of these models, like its one dimensional
counterpart, is a homogeneous function of the coordinates with degree -2.

The plan of the paper is as follows.
In Sec. II, the mapping between the SHO and the Coulomb-like CSM 
problems is established through an underlying $SU(1,1)$ algebra which is
shown to exist in both
the problems. In particular, in Sec. II.A, we discuss the
underlying $SU(1,1)$ algebra in the CSM with the Coulomb like potential. In
Sec. II.B, similar algebraic structure of the many-body systems with the SHO 
potential is presented. The mapping between the two 
is established in Sec. II.C.
In Sec. III, we discuss
the explicit coordinate transformation which maps one problem on to the
other. We find a set of coupled second order nonlinear differential equations,
the solution of which determines the explicit form of the coordinate
transformation. We also solve this differential equation for some specific
many-particle systems.
Discussions have been made in Sec. IV regarding the higher dimensional
generalization of the mapping relating these two type of Hamiltonians.
Finally, in Sec. V, we summarize the results obtained in this paper 
and point out some of the open problems.
In appendix A, we present the energy spectrum and some of the eigen functions
of the Coulomb-like CSM of $B_N$ type. In Appendix B, we show that 
the Casimir operator of the $SU(1,1)$ group 
is the angular part of the CSM Hamiltonian
corresponding to the Coulomb-like or the 
oscillator problems. We also indicate here 
how the group property enables us to use the method of separation of
variables.

\section{The mapping}

\subsection{Algebra of the Coulomb-like problem}

Let us consider the Hamiltonian ($\hbar=m=1$),
\be
H_C = -\frac{1}{2} \bigtriangleup_x + V(x_1,...,x_N)
- \frac{\alpha}{x},
\label{eq0}
\ee
\noindent where,
\be
x=\sqrt{\sum_{i=1}^N x_i^2},  \ \ \ \
\bigtriangleup_x=\sum_{i=1}^N \frac{\partial^2}{\partial x_i^2}.
\label{eq2}
\ee
\noindent
The coordinates of the $N$ particles are denoted by $x_i$ in (\ref{eq2}).
We fix the convention that all Roman indices run from $1$ to $N$ while
all Greek indices run from $1$ to $N^\prime$.
The many-body interaction $V(x)$ in (\ref{eq0}) is 
homogeneous function of degree $-2$.
In particular,
\be
\sum_{i=1}^N x_i  \frac{\partial V}{\partial x_i} = -2 V.
\label{eq1}
\ee
\noindent
It may be noted that the potential term $V$ of the rational CSM of $A_n$ type,
\be
V_{A_n}(\{x_i\}) = \frac{g}{2} \sum_{i<j} (x_i-x_j)^{-2},
\label{eq1.0}
\ee
\noindent indeed satisfies this condition.
In fact the long range interaction terms of the rational $BC_N, B_N, C_N, D_N$
type CSM also 
satisfy this condition. In particular,
\be
V_{BC_N}(g_1,g_2,g_3) = \frac{g_1}{2} \sum_{i<j} \left [ (x_i-x_j)^{-2} +
(x_i+x_j)^{-2} \right ]
+g_2 \sum_i x_i^{-2} + \frac{g_3}{2} \sum_i x_i^{-2},
\label{eq1.1}
\ee
\noindent $V_{B_n} (=V_{BC_n}(g_1,g_2,g_3=0)$), 
$V_{C_n} (=V_{BC_n}(g_1,g_2=0,g_3)$)
and
$V_{D_n} (=V_{BC_n}(g_1,g_2=0, g_3=0)$) have the property (\ref{eq1}). 
Unless mentioned otherwise,  throughout this paper we consider 
arbitrary $V(x)$ satisfying the property (\ref{eq1})
even though schematically we write it as
$V(x)$. 

Let us define the operators $k_1, k_2, k_3$ as,
\bea
& & k_1 = \frac{1}{2} \left ( x \bigtriangleup_x - 2 x V(x) +
x \right ),\nonumber \\
& & k_2=i \left ( \frac{N-1}{2} + \sum_i x_i \frac{\partial}{\partial x_i}
\right ),\nonumber \\
& & k_3 = -\frac{1}{2} \left ( x \bigtriangleup_x - 2 x V(x)
- x \right ). 
\label{eq1.11}
\eea
\noindent It is easily shown that 
these three operators constitute a $SU(1,1)$ algebra, namely,
\be
[ k_1, k_2]= -i k_3, \ \ \ [ k_2, k_3] = i k_1, \ \ \ [k_3, k_1] = i k_2
\, .
\label{eq1.2}
\ee
\noindent Let us emphasis again that the 
$SU(1,1)$ algebra as given here in terms of
the generators $k_1$, $k_2$ and $k_3$ is 
valid for any $V$ satisfying Eq.
(\ref{eq1}). We now show that the eigenvalue equation 
for the Hamiltonian as given by Eq. (\ref{eq0}) can also be
written as an
eigenvalue equation for 
the generator of $SU(1,1)$. To see this, note the following identity,
\be
(k_1+k_3) H_C = - \frac{1}{2} (k_1-k_3) - \alpha \, .
\label{eq3}
\ee
\noindent  Now, following the standard procedure \cite{barut} 
and with the help of
Eq. (\ref{eq3}), the eigenvalue equation,
\be
H_C |N,M> = E_M |N,M> \, ,
\label{eq4}
\ee
\noindent  can be written as,
\be
\left [ k_3 - \frac{\alpha}{\sqrt{-2 E_M}} \right ]
e^{i k_2 \theta_M} |N,M>=0,
\label{eq5}
\ee
\noindent where the function $\theta_M$ is defined by,
\be
\cosh \theta_M = \frac{1 - 2 E_M}{\sqrt{-8 E_M}}, \ \ \
\sinh \theta_M = - \frac{1 + 2 E_M}{\sqrt{-8 E_M}}.
\label{eq6}
\ee
\noindent Thus, the eigenvalue equation for $H_c$ has been transformed into an
eigenvalue equation for the generator $k_3$. The eigen vector
$|N,M>$ with the eigen value $E_M$
in (\ref{eq4}) is defined to characterize the $N$ particle state
with $M$ as the principal quantum number. In general, $M$ can be
expressed as a sum of different non-negative integers
to characterize the degenerate states, depending on the
the particular form of $V(x)$. Even though we do not address here
the question of degeneracy of the many-body system, it should be
noted that the eigen vectors $|N,M>$ do not span the whole eigen
space of $H_c$. In particular, the eigen states $|N,M>$ transform
under the unitary irreducible representations of $SU(1,1)$
labeled by a real constant $\phi(<0)$, where $\phi$ is
related to the eigen value $q$ of the Casimir operator as,
$q=\phi (\phi+1)$. Thus, $|N,M>$ belongs to the $SU(1,1)$ orbit
of the ground state $|N,M;\phi=\phi_0>$, where $\phi_0$ denotes the
minimum admissible value of $\phi$.
As shown in the appendix B, the energy eigen value $E_M$ is
determined in terms of the Casimir operator of $SU(1,1)$ as
\be
E_{m,q}= -\frac{\alpha^2}{2} \left [  m + \frac{1}{2} +
(q+\frac{1}{4})^{\frac{1}{2}} \right ]^{-2},
\label{eq6.e}
\ee
\noindent where $m$ is a nonnegative integer and $q$ is the
eigenvalue of the Casimir operator.

\subsection{Algebra of the oscillator problem}

Let us consider the Hamiltonian ($\hbar=m=1$),
\be
H_{sho}= \frac{1}{2} \left (-\bigtriangleup_y + y^2 + 2 V(y) \right ),
\label{eq7}
\ee
\noindent where,
\be
\bigtriangleup_y=\sum_{\mu=1}^{N^\prime}
\frac{\partial^2}{\partial y_{\mu}^2}, \ \ \
y^2=\sum_{\mu=1}^{N^\prime} y_{\mu}^2.
\label{eq8}
\ee
\noindent The potential $V(y)$ is again homogeneous function of y 
with degree $-2$, 
i.e. it satisfies a
condition analogous to Eq.
(\ref{eq1}). 

We now define three operators $k_1$, $k_2$ and $k_3$ for the oscillator 
as follows,
\bea
& & k_1=\frac{1}{4} \left ( \bigtriangleup_y + y^2 - 
2 V(y) \right ),\nonumber \\ 
& & k_2=\frac{i}{4} \left ( N^\prime + 2 \sum_\mu y_\mu 
\frac{\partial}{\partial y_\mu} \right ),\nonumber \\
& & k_3=\frac{1}{2} H_{sho}.
\label{eq9}
\eea
\noindent Note that these three operators again constitute a $SU(1,1)$ algebra
and the Hamiltonian is proportional to $k_3$. As a result, the eigenvalue
 equation
of the Hamiltonian is also the eigenvalue equation for the operator $k_3$. In
particular,
\be
H_{sho} |N^\prime, M^\prime> = e_{M^\prime} |N^\prime, M^\prime> \ \
\rightarrow \ \ 
k_3 |N^\prime, M^\prime> = \frac{1}{2} e_{M^\prime} |N^\prime, M^\prime>.
\label{eq10}
\ee
\noindent The eigen vector $|N^\prime, M^\prime>$ with the
eigen value $e_{M^\prime}$ in Eq. (\ref{eq10})
is defined, as in the case of the Coulomb problem in Sec.
II.A, to characterize the $N^\prime$
particle state with $M^\prime$ as the principal quantum number.
The eigen states $|N^\prime, M^\prime>$ transform under the unitary
irreducible representations of $SU(1,1)$, labeled by a real 
constant $\phi^\prime(<0)$, where $\phi^\prime$
is related to the eigen value $q$ of the Casimir operator as,
$q=\phi^{\prime} (\phi^{\prime}+1)$.  Thus, $|N^\prime, M^\prime>$
do not span the whole eigen space of $H_{sho}$. Instead, it
belongs to the $SU(1,1)$ orbit of the ground state
$|N^\prime,M^\prime; \phi^\prime=\phi_0^\prime>$, where 
$\phi_0^\prime$ denotes the minimum admissible value of
$\phi^\prime$.
We do not address the question of degeneracy in this paper.
In appendix B, we again show that the energy eigen value 
$e_{M^\prime}$ is determined  in terms of the Casimir operator of 
$SU(1,1)$ as 
\be
e_{m^\prime,q}= 2 m^\prime + 1 + (1+4 q)^{\frac{1}{2}},
\label{eq10.e}
\ee
\noindent where $m^\prime$ is a nonnegative integer and $q$ is the eigen value
of the Casimir operator.
We show in Appendix B that different
representations of the Casimir
operator in terms of the generators (\ref{eq1.11}) and (\ref{eq9}) correspond
to the angular part of $H_c$ and $H_{sho}$ respectively (apart from
a constant).

\subsection{The relationship}

In order to obtain the relationship between the eigen-spectrum of the two CSM
problems, we assume that
the potentials $V(x)$ and $V(y)$  
have the same functional dependence
on the $x$ and the $y$ coordinates respectively. However, the strength of the
interaction may be different in the two cases which we do not mention here
explicitly in order to avoid notational clumsiness.

We have considered two different representations for the generators of the
$SU(1,1)$ algebra, given by (\ref{eq1.11}) and (\ref{eq9}). However, in both 
cases one is using the same positive discrete series
representation of the $SU(1,1)$ algebra. 
Further, in this representation, $k_3$ is taken to be diagonal in
both the cases.
Thus, the isomorphism between the two sets of eigenvectors corresponding to two different
representations of the generators of $SU(1,1)$ naturally follows.
Now note that
both Eqs. (\ref{eq5}) and (\ref{eq10}) are eigenvalue equation
for $k_3$.
Thus, on comparing these two
equations, we have,
\be
|N^\prime, M^\prime> = e^{i k_2 \theta_M} |N, M>, \ \ \
e_{M^\prime} = \frac{\sqrt{2} \alpha}{\sqrt{- E_M}},
\label{eq11}
\ee
\noindent or,
\be
|N,M> = e^{-i k_2 \theta_M} |N^\prime, M^\prime>, \ \ \
E_M= - \frac{2 \alpha^2}{(e_{M^\prime})^2} .
\label{eq12}
\ee
\noindent This establishes the mapping between the eigenvalues as
well as the eigenfunctions of $H_c$ and $H_{sho}$.
This also implies that $H_c$ is exactly solvable provided $H_{sho}$
is so and vice versa. 
Since this analysis is valid for any $V(x)$ satisfying Eq. (\ref{eq1}), 
this means that we have found a class of new, exactly solvable, many-body 
problems in one dimension.
For example, the $B_N, C_N, D_N, BC_N$
CSM, with the harmonic oscillator potential replaced by the Coulomb-like
potential must also be exactly solvable many-body problems. 
As an illustration,  the eigenvalues as well as some of the eigen
functions of the $B_N$-model with Coulomb-like potential have been 
worked out in Appendix A.

The second relation in Eq. (\ref{eq11}) as well as (\ref{eq12}) describes the
relationship between the energy spectra of the two problems. 
The fact that this relationship is indeed valid is easily
checked by using Eqs.
(\ref{eq6.e}) and (\ref{eq10.e})
and identifying $m$ as $m^\prime$. 
Since, the eigen value $q$ of the Casimir operator is independent
of any particular representation of the generators (i.e. Eq. (\ref{eq1.11})
or (\ref{eq9}) ), it is expected
that the comparison of the known energy spectra
of $H_c$ and $H_{sho}$
would in general relate different quantum numbers as well as parameters
of a particular theory to the another. We work
out here some known examples to explore such relations.\\
(a) Let us first consider
a simple example i.e. consider the potentials,
\be
V(x)= g x^{-2}, \ \ \ V(y)=g^\prime y^{-2} \, .
\ee
\noindent The energy eigen values $E_m$ and $e_m^\prime$ for this choice
of $V(x)$ and $V(y)$ are given by,
\be
E_{m,k}=-\frac{\alpha^2}{2} \left ( m + \frac{1}{2} + \lambda_k \right )^{-2},
\ \ \
e_{m^\prime,k^\prime} = 2 m^\prime + 1 + \lambda_{k^\prime}^\prime,
\label{ener}
\ee
\noindent where $\lambda_k$ and $\lambda_{k^\prime}^\prime$ are defined as,
\be
\lambda_k = \left [ \frac{1}{2} ( 2 k + N -2 ) + 2 g \right ]^{\frac{1}{2}},
\ \ \
\lambda_{k^\prime}^\prime = \left [ \frac{1}{2} (2 k^\prime + N^\prime -2)
+ 2 g^\prime \right ]^{\frac{1}{2}}.
\label{ener1}
\ee
\noindent One can easily see that Eqs. (\ref{eq12}), (\ref{ener})
and (\ref{ener1}) are consistent with each other 
provided the following relations hold
good
\be
N^\prime = 2 (N-1), \ \ \ g^\prime=\frac{g}{4}, \ \ \ k^\prime = 2 k, \ \
\ m^\prime = m \, .
\label{ct}
\ee
\noindent We will see in the next section that the 
first two relations also follow from the coordinate transformation.

(b) Consider the rational CSM (with SHO) of $A_n$ type and the corresponding
Coulomb-like problem \cite{iop}. In this case, 
the energy eigenvalues $E_{m,k}$ and $e_{m^\prime,k^\prime}$
are given by,
\be
E_{m,k}=-\frac{\alpha^2}{2} \left [ m + k + b + \frac{1}{2} \right ]^{-2}, \ \ \
e_{m^\prime, k^\prime} = 2 m^\prime + k^\prime + b^\prime + 1,
\label{fe}
\ee
\noindent where $2 b= (N-1) (1+\lambda N)-1$, $2 b^\prime=
(N^\prime-1) (1+\lambda^\prime N^\prime) -1$, $g=\lambda (\lambda-1)$ and
$g^\prime=\lambda^\prime (\lambda^\prime-1)$. Now observe that 
Eqs. (\ref{eq12}) and
(\ref{fe}) are consistent with each other 
provided the first, the third and the fourth
relations of Eq. (\ref{ct}) are valid and further the following
relation between $\lambda$ and $\lambda^\prime$ holds true,
\be
\lambda^\prime = \frac{N}{2 N-3} \lambda \, .
\label{fe1}
\ee

(c) Finally consider the rational CSM of $B_n$ type and the corresponding
Coulomb-like problem (See Appendix A). The energy eigenvalues
$E_{m,k}$ and
$e_{m^\prime,k^\prime}$ corresponding to these two cases are,
\be
E_{m,k}=-\frac{\alpha^2}{2} \left [ m + 2 k + b +
\frac{1}{2} \right ]^{-2}, \ \
e_{m^\prime, k^\prime}= 2 (m^\prime + k^\prime) + b^\prime + 1,
\label{comp}
\ee
\noindent where $2 b=(N-1) (1+2 \lambda N) +2 \lambda_1 N -1$ and
$2b^\prime=(N^\prime-1) (1+2 \lambda^\prime N^\prime) +
2 \lambda_1^\prime N^\prime-1$.
Again, it follows
that Eqs. (\ref{eq12}) and (\ref{comp}) are consistent with each other 
provided the first, third and the fourth relations of Eq. (\ref{ct}) are 
valid and further, the
following relation among $\lambda$'s holds true,
\be
\lambda_1^\prime + (2N-3) \lambda^\prime -
\lambda_1 \frac{N}{N-1} -\lambda N=0.
\label{la}
\ee
\noindent Note that Eq. (\ref{la})
is satisfied provided 
$\lambda_1^\prime=\frac{N}{N-1} \lambda_1$ and 
$\lambda$ and $\lambda^\prime$ are related as in the previous case i.e.
by Eq. (\ref{fe1}). 
It may be noted here that for the $D_N$ case $\lambda_1 =
\lambda_1^\prime=0$,
and hence in that case the relation (\ref{la}) reduces to (\ref{fe1}).  

Summarizing, we find that for all types of CSM models 
in one dimension, 
the mapping between the oscillator and the Coulomb-like $N$-body problems
holds good provided the first, third and the last relations of Eq.
(\ref{ct}) are valid.
It is worth pointing out that the first relation of Eq.
(\ref{ct}) is also dictated by the coordinate transformation and is independent
of the particular from of $V(x)$, as will be seen in the
next section. 
It is amusing to note that the third and the 
fourth relations of Eq. (\ref{ct}) are also true for the
usual SHO and the Coulomb problems \cite{china2}. Thus, these must be universal
relations valid  for any $V(x)$ since these relations are also valid 
in the limit of vanishing
$V(x)$. 
Note however that 
the relation between $\lambda$ and $\lambda^\prime$ is
dependent on the particular form of $V(x)$.
Finally, it seems that relation (\ref{fe1}) is universal
in some sense for the mapping between the rational CSM of all types and 
the corresponding Coulomb-like problems.

\section{Coordinate transformation}

In this section, we will be discussing about the explicit coordinate 
transformation relating CSM with the oscillator and the Coulomb-like 
potentials. 
On comparing Eqs. (\ref{eq1.11}) and (\ref{eq9}), we have the
following operator relations
\be
x=\frac{1}{2} y^2,
\label{eq13}
\ee
\be
\frac{N-1}{2} + \sum_i x_i \frac{\partial}{\partial x_i} =
\frac{1}{4} \left [ N^\prime +
2 \sum_\mu y_\mu \frac{\partial}{\partial y_\mu} \right ],
\label{eq14}
\ee
\be
x \bigtriangleup_x - 2 x V(\{x_i\}) = \frac{1}{2} \left [ \bigtriangleup_y
- 2 V(\{y_\mu\}) \right ].
\label{eq15}
\ee
\noindent Let us now assume a coordinate transformation of the form
\be
x_i = f_i(\{y_\mu\}), 
\label{eq16}
\ee
\noindent where $f_i$'s are $N$ arbitrary functions of the
coordinates $y_\mu$'s
with the constraint $\sqrt{\sum_i f_i^2}=\frac{1}{2} y^2$.
The particular form as well as the properties of all the $f_j$'s will be
determined from Eqs. (\ref{eq13}) to (\ref{eq15}). On multiplying
both sides of Eq. (\ref{eq14})  by $x_j$ from right and using 
relation
(\ref{eq16}), we encounter two
different cases.\\
(a) 
\be 
N^\prime = 2 (N+1), \ \ \ \ \sum_\mu y_\mu \frac{\partial f_i}{\partial y_\mu}
= 0.
\label{eq17}
\ee
\noindent However, the second relation of Eq. (\ref{eq17}) implies that all
$f_i$'s are homogeneous function of degree zero which is in direct
contradiction with Eq. (\ref{eq13}). Thus, this possibility is ruled out.\\
(b)
\be
N^\prime = 2 ( N + 1 -d ), \ \ \
\sum_\mu y_\mu \frac{\partial f_i}{\partial y_\mu}= d f_i.
\label{eq18}
\ee
\noindent The second relation of Eq. (\ref{eq18}) implies that all $f_i$'s
are homogeneous function of degree $d$. However, it follows from 
Eq. (\ref{eq13})
that  $d$ must be 2 and hence the first relation of Eq. (\ref{eq18}) now reads
as,
\be
N^\prime = 2 ( N -1).
\label{eq19}
\ee
\noindent Equation (\ref{eq19}) establishes a relationship between the
total number of particles in the two cases. 
Notice that Eq. (\ref{eq19}) also followed from a comparison of the 
eigenvalues in the two cases (see Eq. (\ref{ct}).
It is amusing to note that 
exactly the same relation is also obtained in case
one considers the mapping between the usual $N$ dimensional Coulomb
and $N^\prime$ dimensional harmonic oscillator problems. In other words,
(\ref{eq19}) is independent of the particular form of the many-particle
potential.

On multiplying 
both sides of Eq. (\ref{eq15}) 
by $x_j$ from right and  
using relation (\ref{eq16})
we obtain,
\be
\left [ \bigtriangleup_y - 2 V(\{y_\mu\}) \right ] f_i(\{y_\mu\})
+ 2 y^2 V(\{f_j\}) f_i(\{y_\mu\})=0 \, .
\label{eq20}
\ee
\noindent This is a set of highly nonlinear second order differential
equation. Moreover, only those solutions for which all $f_i$'s are
homogeneous function of degree 2 and the norm of $f_i$'s is $
\frac{1}{\sqrt{2}} y$
are acceptable solutions for our purpose.

One would now like to
ask if such a solution (to Eq.
(\ref{eq20})) exists or not. Note at this point that for acceptable solutions, 
the first term of (\ref{eq20}) (i.e. $L_i=
\bigtriangleup_y f_i$) should either be a constant or be a homogeneous
function of degree zero. Let us first consider the case $L_i=0$,
i. e., those solutions
which are also solutions
of $N^\prime$ dimensional Laplace equation. With the use of (\ref{eq20}),
this implies following relation between $V(x)$ and $V(y)$,
\be
V(x)=y^{-2} V(y).
\label{tt}
\ee
\noindent In that case the operator relations (\ref{eq13}), (\ref{eq14}) 
and (\ref{eq15})
are identical to those in the case of the  usual $H_{sho}$ and $H_c$ problems. 
Now exactly following the procedure as given in Zeng et al. \cite{china2}, we
find one valid coordinate transformation between the two problems as
given by
\be
x_i=f_i= \frac{1}{4} \sum_{\alpha, \beta}
\Gamma_{\alpha \beta}^i y_\alpha y_\beta,
\label{tt0}
\ee
\noindent where the matrices
$\Gamma$ constitute the Clifford algebra,
\be
\Gamma^i \Gamma^j + \Gamma^j \Gamma^i = 2 \delta^{ij}.
\label{tt1}
\ee
\noindent We might add here that the coordinate transformation (\ref{tt0})
can be written down explicitly with the use of the real representation of the
Clifford algebra \cite{ritten}. However, we have to determine the form
of $V(x)$ such that Eq. (\ref{tt}) is consistent with the coordinate
transformations as given by (\ref{tt0}) and (\ref{tt1}).
One such choice is, $V(x)= 4 g x^{-2}$ and $V(y)=g y^{-2}$.
We may add here that unfortunately, none of the inverse square 
interactions of the CSM satisfy
(\ref{tt}).

Let us now consider the second possibility i.e. 
all $L_i$'s are nonzero arbitrary constants. In this case $x_i$'s are not
independent of each other and no valid solution can be found. Thus,
it seems that $\L_i$'s as homogeneous functions of degree zero
is probably the only  alternative for finding explicit coordinate
transformation
in the interesting case of CSM.
However, finding such solutions explicitly or even proving the
existence of such
solutions is a highly nontrivial problem and at present we do not have any
answer to this question.

Finally, as an aside, let us note that 
the mapping between $H_c$ and $H_{sho}$
as described by Eqs.
(\ref{eq11}) and (\ref{eq12}) is valid even when the many-body interaction
of the two problems is not the same (i.e. they have completely different
functional dependence ). 
However, both should satisfy the homogeneity 
condition (\ref{eq1}). In such cases let us denote $V(y)$ by $\tilde{V}
(y)$. Now note that the coordinate transformation (\ref{tt0}) can be identified
as the required coordinate transformation provided it relates $V(x)$
and $\tilde{V}(y)$ as follows,
\be
V(x)= y^{-2} \tilde{V}(y), \ \ \ \tilde{V}(y)= 2 x V(x).
\label{re}
\ee
\noindent Thus, with each type of rational CSM one can associate
a new many-body problem with Coulomb-like interaction which are related
by the coordinate transformation (\ref{tt0}). Similarly, one can
find new many-body Hamiltonians with oscillator confinement associated
with $H_c$. In particular, $H_c$ with $V(x)$ given by (\ref{eq1.0})
is related to $H_{sho}$ with $\tilde{V}(y)$ given by,
\be
\tilde{V}(y) =8 g y^2 \sum_{i<j} \left [ \sum_{\alpha \beta}
\left ( \Gamma^i-
\Gamma^j \right )_{\alpha \beta}  y_\alpha y_\beta \right ]^{-2}, 
\label{eq}
\ee
\noindent where $(\Gamma_i-\Gamma_j )_{\alpha \beta}$ implies $\alpha \beta$
element of the matrix $\Gamma_i-\Gamma_j$. Note that for the real
representation
of the Clifford algebra \cite{ritten}, some of the $\Gamma$'s are diagonal
and, consequently, the many-body interaction (\ref{eq}) is
$N^\prime(=2 (N-1))$-
body interaction unlike in the case of usual CSM. This type of new many-body
Hamiltonians may or may not be
interesting
from the physical point of view. However, they have the remarkable
property of being exactly solvable.

\section{The mapping : higher dimensional generalization}

In the last two sections we have established the mapping between 
the oscillator and the Coulomb-like
problem in one dimensional many-body systems.
We now generalize these results to higher dimensional many-body systems
with the many-body interactions as homogeneous functions of degree $-2$.
Recall at this point that the many-body interaction of all the known
higher dimensional CSM type models is homogeneous with degree $-2$. For
example, the Calogero-Marchioro model \cite{marchioro}, models with novel
correlations \cite{nc}, models with two-body interactions \cite{me} and
models considered in \cite{gamb,pani}
have this property. 

Let us consider the operators $K_1$, $K_2$ and $K_3$
for the Coulomb-like problem as follows,
\bea
& & K_1 = \frac{1}{2} \left ( X \bigtriangleup_X - 2 X V(X) +
X \right ),\nonumber \\
& & K_2=i \left ( \frac{N D-1}{2} + \sum_i \vec{r}_i . \vec{\bigtriangledown}_i
\right ),\nonumber \\
& & K_3 = -\frac{1}{2} \left ( X \bigtriangleup_X - 2 X V(X)
- X \right ), 
\label{ma1}
\eea
\noindent where,
\be
X=\sqrt{\sum r_i^2}, \ \ \ \bigtriangleup_X=\sum_i \bigtriangledown_i^2,
\label{ma2}
\ee
\noindent and $\vec{\bigtriangledown}_i$ is the $D$ dimensional
gradient of the $i$th particle.
The potential $V(X)$ is homogeneous with degree $-2$
and satisfies the homogeneity condition analogous to equation
(\ref{eq1}). One can check that these three operators constitute the $SU(1,1)$
algebra (\ref{eq1.2}). The eigen equation of the Hamiltonian,
\be
H_c^D= - \frac{1}{2} \bigtriangleup_X + V(X) - \frac{\alpha}{X}
\label{ma3}
\ee
\noindent can be shown to be given by equation (\ref{eq5}) with $k_2$ and $k_3$
replaced by $K_2$ and $K_3$ respectively.

Similar to the one dimensional oscillator problem, we define the three
operators
for the $D^\prime$ dimensional many-body problem with oscillator 
potential as \cite{gamb},
\bea
& & K_1=\frac{1}{4} \left ( \bigtriangleup_Y + Y^2 - 
2 V(Y) \right ),\nonumber \\ 
& & K_2=\frac{i}{4} \left ( N^\prime D^\prime + 2 \sum_\mu \vec{r}_\mu^\prime 
. \vec{\bigtriangledown}_\mu^\prime \right ),\nonumber \\
& & K_3=\frac{1}{2} H_{sho}^{D^\prime} = \frac{1}{4} \left ( - 
\bigtriangleup_Y + Y^2
+ 2 V(Y) \right ),
\label{ma4}
\eea
\noindent where $\bigtriangleup_Y$ and $Y^2$ are given as,
\be
\bigtriangleup_Y = \sum_\mu {\bigtriangledown_\mu^\prime}^2, \ \ \
Y=\sqrt{\sum_\mu {r_i^\prime}^2}.
\label{ma5}
\ee
\noindent We denote $\bigtriangledown_\mu^\prime$ as the $D^\prime$ dimensional
gradient operator for the $\mu$th particle. These three operators satisfy the
$SU(1,1)$ algebra (\ref{eq1.2}) and the Hamiltonian is proportional to $K_3$.

Following the discussions of Sec. II.C, one can establish the mapping between
the eigen values as well as the eigen vector of $H_c^D$ and the same quantities
of $H_{sho}^{D^\prime}$. Equations (\ref{eq11}) and (\ref{eq12}) continue to 
be valid in  
the higher dimensional case also but with 
$k_2$ replaced by $K_2$. In particular,
\be
|N,D,M> = e^{-i K_2 \theta_M} |N^\prime, D^\prime, M^\prime>, \ \ \ \
E_M = - \frac{2 \alpha^2}{(e_{M^\prime})^2}.
\ee
\noindent An analysis of Eqs. (\ref{ma1}), (\ref{ma2})
, (\ref{ma4}) and (\ref{ma5}), on the lines of  
what has been done in the previous section, shows that the relation,
\be
N^\prime D^\prime = 2 (N D -1),
\ee
\noindent holds true for any $V(X)$. Note that this equation reduces to
(\ref{eq19}) for $D=D^\prime=1$.

We would like to emphasize here that unlike the one dimensional case,  
the higher dimensional many-body systems, like Calogero-Marchioro
model \cite{marchioro} or the
models for novel correlations \cite{nc}, have a part of 
the energy spectrum with a linear dependence and the remaining part with a  
nonlinear dependence
on the coupling constant of the relevant problem. Unfortunately, so far, 
only the linear part
of the spectrum has been 
obtained analytically for all the known higher dimensional many-body problem.
In fact, not even one energy level with nonlinear dependence on the coupling
constant has been obtained as yet. Not surprisingly,  
even using the underlying $SU(1,1)$ symmetry of the Calogero-Marchioro problem,
one can not find the missing non-linear part \cite{gamb}. This is because the
angular part of the Hamiltonian or equivalently the eigenvalue problem of
the Casimir operator can not be solved exactly in higher dimensions. Thus,
we are unable to compare the energy spectra of $H_c^D$ and $H_{sho}^{D^\prime}$ in
higher dimensions as has been done for the one dimensional systems. 

\section{Summary}

In this paper 
we have shown that the energy spectrum as well as the eigen functions of the
rational CSM with Coulomb-like interaction 
associated with the root structure of $A_N$, $B_N$, $C_N$, $D_N$
and
$BC_N$ can
be obtained from the corresponding CSM with the harmonic oscillator
potential. 
Consequently, all types of CSM with a Coulomb-like interaction are
also exactly solvable models. Thus, one has added a new class of members to the
family of exactly solvable many-body systems in one dimension. 
Further, we
have shown that all these results can be generalized to other many-body systems
in one dimension
provided the many-particle interaction of these systems, much akin to the
CSM, is a homogeneous function
of degree -2. We have explicitly found the coordinate transformation for some
specific cases which maps the Coulomb-like problem to a harmonic
one. Though we are not able to find the coordinate transformation
responsible for such mapping for each and every case, we have found 
a set of second order coupled nonlinear differential equation and show
that a particular class of solutions of this set of equations are going
to determine the coordinate transformation. However, the proof of
existence of such
class of solutions and, if possible, to find them explicitly is a
highly nontrivial problem. 

\acknowledgements{One of us (PKG) would like to thank Institute of Physics,
Bhubaneswar, where part of the work has been carried out.}

\begin{appendix}

\section{$B_N$ CSM with Coulomb-like potential} 

In this Appendix we obtain the spectrum as well as the eigen-functions of the 
$B_n$ type CSM with Coulomb-like potential.
In particular, we consider the Hamiltonian
(\ref{eq0}) with $V(x)$ given by (\ref{eq1.1}) and
$g_1=\lambda (\lambda-1)$, $g_2=\lambda_1 (\lambda_1-1)$ and $g_3=0$. Note that
the energy eigen states of $BC_N$ as well as $C_N$ CSM could be obtained easily
from the known results of $B_N$ CSM. Let
\be
\Phi= \prod_l x_l^{\lambda_1} \prod_{i <j} (x_i^2 -
x_j^2)^\lambda P_{2k}(x) \phi(x)
\label{a1}
\ee
\noindent be a solution of the Schr$\ddot{o}$dinger equation
$H_c \Phi=E \Phi$.
In Eq. (\ref{a1}), $P_{2k}(x)$ is a symmetric homogeneous polynomial
of the coordinates with degree $2 k$
and satisfies the generalized Laplace equation,
\be
\bigtriangleup_x P_{2k}(x) +
2 \lambda_1 \sum_i x_i^{-1} \frac{\partial P_{2k}}{\partial x_i}
+ 4 \lambda \sum_{i \neq j}
\frac{x_i}{x_i^2-x_j^2} \frac{\partial P_{2k}}{\partial x_i}=0.
\label{a2}
\ee
\noindent Plugging the expression (\ref{a1}) into the Schr$\ddot{o}$dinger
equation, we have,
\be
\phi^{\prime \prime} + \left [ 2 b + 4 k + 1 \right]
\frac{\phi^\prime}{x} + 2 \left ( E + \frac{\alpha}{x} \right ) \phi =0,
\label{a3}
\ee
\noindent where the parameter $b$ is given by,
\be
b=\frac{1}{2} (N-1) (1+2 \lambda N) - \frac{1}{2} + \lambda_1 N.
\ee
\noindent Defining a new variable $t=\sqrt{2 E} x$,
Eq. (\ref{a3})
can be solved as,
\be
\phi_{n,k}=exp(-t) L_n^{2 b + 4 k}(2 t),
\label{a4}
\ee
\noindent where $L_n^{2 b +4 k}(2 t)$ is the Laguerre polynomial with the
argument $ 2 t$. The energy eigen values corresponding to the wave functions 
(\ref{a1}) are,
\be
E_{n,k}= -\frac{\alpha^2}{2} \left [n + 2 k + \frac{1}{2} + b \right ]^{-2}.
\label{a5}
\ee
\noindent It may be noted here that the results for the $D_N$ case 
can be obtained from here simply by putting $\lambda_1=0$.

The wave function given by (\ref{a1}) contains a homogeneous function
$P_{2k}$ of degree $2 k$ which is determined by Eq. (\ref{a2}).
In general, we do not know the exact solutions of Eq. (\ref{a2}) except for
some small values of $N$ and $k$. However, it can be shown that 
Eq. (\ref{a2}) is exactly solvable by following the methods
described in Brink et al. \cite{turb}. 
In fact, apart from some constant, the corresponding
equation in \cite{turb}
contains one more extra term $\sum_i x_i \frac{\partial}{\partial x_i}$
than (\ref{a2}) and the treatment as well as conclusions obtained there 
are also valid in the case of Eq. (\ref{a2}).

\section{Casimir Operator and separation of variables}

In this Appendix we study the role of the Casimir 
operator of the $SU(1,1)$ group regarding
the separation of variables in the case of Schr$\ddot{o}$dinger equation for 
$H_c$ and $H_{sho}$. The Casimir operator
of $SU(1,1)$ for the class of unitary irreducible representations,
called the positive
discrete series, is defined by \cite{barut,gamb}
\be
C=k_3^2 - k_1^2 - k_2^2,
\label{b2}
\ee
\noindent and it commutes with all the generators $k_1$, $k_2$ and $k_3$.
The Casimir operator and $k_3$ are diagonal in this representation and
the eigenvalue of $k_3$ is given by,
\be
\epsilon_{\pm}= n + \frac{1}{2} \pm \left (q+
\frac{1}{4} \right )^{\frac{1}{2}},
\label{b2.1}
\ee
\noindent where $n$ is a  nonnegative integer and $q$ is the eigen value of
the Casimir operator. $\epsilon_-$ has the restriction
$(q+\frac{1}{4})^{\frac{1}{2}}
< \frac{1}{2}$ and it leads to physically unacceptable
solutions \cite{gamb}. Thus,
we will be concerned with $\epsilon_+$ only in this paper.

We use the notation $C_N^x$ and $C_{N^\prime}^y$ for the Casimir
operators associated with the generators of the $SU(1,1)$ given by
two different representations (\ref{eq1.11}) and (\ref{eq9}) respectively.
Plugging (\ref{eq1.11}) and (\ref{eq9}) into (\ref{b2}) and after some
manipulation \cite{gamb}, we find,
\bea
C_N^x & = & \frac{1}{4} (N-1) (N-3) + 2 x^2 V(x) - \sum_{i<k} \left (
x_i \frac{\partial}{\partial x_k} -
x_k \frac{\partial}{\partial x_i} \right )^2,\nonumber \\
C_{N^\prime}^y & = & \frac{1}{16} N^\prime (N^\prime-4) + \frac{1}{2}
y^2 V(y) - \frac{1}{4} \sum_{\mu < \nu} \left
( y_\nu \frac{\partial}{\partial y_\mu} -
y_\mu \frac{\partial}{\partial y_\nu} \right )^2.
\label{b3}
\eea
\noindent Now since $V(x)$ is homogeneous
with degree $-2$, hence, $x^2 V(x)$ can be expressed
purely in terms of the $N-1$ angular variables in the
$N$-dimensional spherical coordinates. Similarly, $y^2 V(y)$
is determined solely in terms of the $N^\prime-1$ angular variables
of the $N^\prime$ dimensional spherical coordinates. Thus,
apart from a constant factor both $C_{N}^x$ and $C_{N^\prime}^y$
are exactly equivalent to the angular part of the respective Hamiltonians
$H_C$ and $H_{sho}$. In particular, the angular part of the Hamiltonians
$H_c$ and $H_{sho}$ is given by,
\be
H_c^a= C_N^x - \frac{1}{4} (N-1) (N-3), \ \ \
H_{sho}^a=C_{N^\prime}^y- \frac{1}{16} N^\prime (N^\prime-4).
\label{b3.a}
\ee
\noindent Further, the constant factor of $C_N^x$ is related to the
constant factor of $C_{N^\prime}^y$ by (\ref{eq19}).
It may be noted here that the
total angular momentum $L^2$ and ${L^\prime}^2$,
\be
L^2= - \sum_{i<k} \left (
x_i \frac{\partial}{\partial x_k} -
x_k \frac{\partial}{\partial x_i} \right )^2, \ \ \
{L^\prime}^2 =
 - \sum_{\mu < \nu} \left
( y_\nu \frac{\partial}{\partial y_\mu} -
y_\mu \frac{\partial}{\partial y_\nu} \right )^2,
\label{b3.l}
\ee
\noindent of $H_c$ and $H_{sho}$
respectively,
are also related to each other as,
\be
{L^\prime}^2= 4 L^2, \ \ \ l^\prime=2 l,
\label{b3.1}
\ee
\noindent in case relation
(\ref{tt}) is satisfied. In Eq. (\ref{b3.1}), $l$ and $l^\prime$ denote
the eigenvalues of $L$ and $L^\prime$ respectively.
This result is also valid in case one starts with
$\tilde{V}(y)$ instead of $V(y)$ in $H_{sho}$ and relation (\ref{re}) holds
true.

Following Gambardella \cite{gamb}, it is easily seen that
the relation $[C,k_3]=0$ implies,
\be
[H_c^{r}, H_c^{a}]=0, \ \ \ [H_{sho}^r, H_{sho}^a]=0,
\label{b4}
\ee
\noindent where $H_c^r$ and $H_{sho}^r$ are the radial part of the
$N$ dimensional conventional Coulomb problem and the $N^\prime$ dimensional
conventional oscillator
problems respectively.
We have used the
relation,
\be
k_3= \alpha + \frac{x}{2} + x H_c,
\label{b5}
\ee
\noindent in order to derive the first equation of (\ref{b4}). The relations
(\ref{b4}) imply that the method of separation of variables is applicable
to both $H_c$ and $H_{sho}$. This relation for $H_{sho}$ in arbitrary
dimensions was known earlier \cite{gamb}, while we have generalized this result
to the case of $H_c$.

\end{appendix}


\begin{references}

\bibitem{cs} F. Calogero, J. Math. Phys. (N.Y.) {\bf 10} (1969) 2191;
 {\bf 10} (1969) 2197.

\bibitem{cs1} B. Sutherland, J. Math. Phys.(N.Y.) {\bf 12} (1971)
246; {\bf 12} (1971) 251; Phys. Rev. {\bf A 4} (1971) 2019.

\bibitem{pr} M. A. Olshanetsky and A. M. Perelomov, Phys. Rep. {\bf 71}
 (1981) 314; {\bf 94} (1983) 6.

\bibitem{iop} A. Khare, J. Phys. A: Math. Gen. {\bf 29} (1996) L45;
cond-mat/9712132. 

\bibitem{china1} G. Zeng, K. Su and M. Li, Phys. Rev. {\bf A 50} (1994) 4373
and references therein.

\bibitem{china2} G. Zeng, S. Zhou, S. Ao and F. Jiang, J. Phys. A: Math. Gen.
{\bf 30}
(1997) 1775.

\bibitem{barut} Theory of group representations and applications, A. O.
Barut and R. Raczka, World Scientific, Singapore.

\bibitem{ritten} M. D. Crombrugghe and V. Rittenberg, Ann. Phys. {\bf
151} (1983) 99.

\bibitem{marchioro}
F. Calogero and C. Marchioro, J. Math. Phys. (N.Y.) {\bf 14} (1973) 182;
A. Khare and K. Ray,  Phys. Lett. {\bf A230} (1997) 139.

\bibitem{nc} M. V. N. Murthy, R. K. Bhaduri and D. Sen,
Phys. Rev. Lett. {\bf 76} (1996) 4103; R. K. Bhaduri, A. Khare,
J. Law, M. V. N. Murthy and D. Sen, J. Phys. {\bf A30} (1997) 2557;
P. K. Ghosh, cond-mat/9607009.

\bibitem{me} P. K. Ghosh, Phys. Lett. {\bf A229} (1997) 203.

\bibitem{gamb} P. J. Gambardella, J. Math. Phys. (N.Y.) {\bf 16} (1975) 1172.

\bibitem{pani}
N. Gurappa, C. N. Kumar and P. K. Panigrahi,
Mod. Phys. Lett. {\bf A11} (1996) 1737;
R. K. Ghosh
and S. Rao, Phys. Lett. {\bf A238} (1998) 213.

\bibitem{turb} L. Brink, A. Turbiner and N. Wyllard, J. Math. Phys. {\bf 39}
 (1998) 1285; W. R$\ddot{u}$hl and A. Turbiner,
Mod. Phys. Lett. {\bf A10} (1995) 2213.

\end{references}
\end{document}